\documentclass[english,aps,pra,superscriptaddress,amsmath,amsfonts,amssymb,floatfix,nofootinbib]{revtex4}
\usepackage[utf8]{inputenc}
\usepackage{color}
\usepackage{babel}
\usepackage{mathtools}
\usepackage{amsmath}
\usepackage{amssymb}
\usepackage{graphicx}
\usepackage[unicode=true,
 bookmarks=false,
 breaklinks=false,pdfborder={0 0 1},backref=false,colorlinks=false]
 {hyperref}

\makeatletter
\@ifundefined{textcolor}{}
{%
 \definecolor{BLACK}{gray}{0}
 \definecolor{WHITE}{gray}{1}
 \definecolor{RED}{rgb}{1,0,0}
 \definecolor{GREEN}{rgb}{0,1,0}
 \definecolor{BLUE}{rgb}{0,0,1}
 \definecolor{CYAN}{cmyk}{1,0,0,0}
 \definecolor{MAGENTA}{cmyk}{0,1,0,0}
 \definecolor{YELLOW}{cmyk}{0,0,1,0}
}

\@ifundefined{date}{}{\date{}}
\usepackage{enumerate}
\usepackage{epsfig}
\usepackage{sidecap}
\usepackage[normalem]{ulem}
\usepackage{color}
\usepackage{enumerate}
\usepackage{bm}

\usepackage{amsfonts}
\usepackage{enumitem}

\usepackage{babel}

\usepackage[caption=false]{subfig}

\@ifundefined{showcaptionsetup}{}{%
 \PassOptionsToPackage{caption=false}{subfig}}
\usepackage{subfig}
\makeatother

\begin{document}
\title{Quasidistribution of phases in Raman process with weak and strong
pumps}
\author{Kishore Thapliyal}
\thanks{Email:kishore.thapliyal@upol.cz}
\affiliation{RCPTM, Joint Laboratory of Optics of Palacky University
and Institute of Physics of Academy of Science of the Czech Republic,
Faculty of Science, Palacky University, 17. listopadu 12, 771 46 Olomouc,
Czech Republic}
\author{Jan Pe$\check{{\rm r}}$ina}
\affiliation{Joint Laboratory of Optics of Palacky University
and Institute of Physics of Academy of Science of the Czech Republic,
Faculty of Science, Palacky University, 17. listopadu 12, 771 46 Olomouc,
Czech Republic}
\affiliation{Department of Optics, Faculty of Science, Palacky University, 17.
listopadu 12, 771 46 Olomouc, Czech Republic}
\begin{abstract}
Nonclassicality is studied through a quasidistribution of phases for
the Raman process under both weak and strong pump conditions. In the
former case, the solution is applicable to both resonant and off-resonant
Raman processes, while strong classical pump is assumed at resonance.
Under weak pump conditions (i.e., in a complete quantum treatment),
the phase difference of phases described by single nonclassical modes
is required to be filtered to describe a regular distribution function,
which is not the case with strong pump. Compound Stokes-phonon mode
shows nonclassical features of phases in both weak and strong pumping,
which effect is similar to that for compound pump-phonon (Stokes-anti-Stokes)
mode with weak (strong) pump. While anti-Stokes-phonon mode is observed
to be classical and coherence conserving in strong pump case, pump-Stokes
mode shows similar behavior in a special case in quantum treatment.
\end{abstract}
\maketitle

\section{Introduction}

Nonclassical states, having negative values of (multimode) Glauber-Sudarshan
$P\left(\left\{ \alpha_{j}\right\} \right)$ $\left(\left\{ \alpha_{j}\right\} =\left\{ \alpha_{1},\alpha_{2}\ldots,\alpha_{n}\right\} \right)$
quasidistribution function \citep{glauber1963coherent,sudarshan1963equivalence},
are significant resource in multifaceted quantum information processing
and technology, namely quantum communication \citep{gisin2007quantum},
computational supremacy \citep{harrow2017quantum}, metrology \citep{giovannetti2011advances},
machine learning \citep{biamonte2017quantum}, sensing \citep{degen2017quantum},
simulation \citep{georgescu2014quantum}, game theory \citep{eisert1999quantum},
etc. Independently, studies related to quantum phase have also garnered
attention over the past few decades (see \citep{perinova1998phase}
for review). Recently, nonclassicality in the phase is also studied
by isolating the role of quantum phase from the classical phase in
optomechanical system \citep{armata2016quantum}. Interestingly, when
having quasidistribution $P\left(\left\{ \alpha_{j}\right\} \right)$
of complex amplitudes $\alpha_{j}=\left|\alpha_{j}\right|\exp\left(i\phi_{j}\right)$,
integrating over the moduli of real amplitudes, we can obtain a quasidistribution
$\Theta\left(\left\{ \phi_{j}\right\} \right)$ of phases $\phi_{j}$.
In the classical region, it is regular, while in the quantum region
it may be singular. However, in \citep{perina2019quasidistribution}
a procedure has been suggested to show that one can obtain a regular
distribution also in the quantum region provided that some phase differences
of the phases are filtered. 

In what follows, we study here the effect of nonclassicality on values
of phase differences allowed to describe the quasidistribution of
phase difference in case of both resonant and off-resonant Raman processes.
Specifically, we illustrate this using the nonlinear optical process
of Raman scattering in finite-time approximation with weak pumping
and in parametric approximation with strong coherent classical pumping.
We mainly analyze conditions for the distribution of phase differences,
whereas their forms are given in \citep{perina2019quasidistribution}.

\section{Quasidistribution of phases in raman process \label{sec:coh}}

We will be considering two scenarios: weak and strong pumps. In the
first case an approximate perturbation solution using a complete quantum
treatment is obtained which is applicable for both resonant and off-resonant
Raman processes \citep{thapliyal2019nonclassicality,thapliyal2019lower}.
In contrast, in the latter case of strong coherent classical pumping,
an exact analytic solution is possible \citep{pieczonkova1981statistical}. 

As the quasidistribution $P\left(\left\{ \alpha_{j}\right\} \right)$
can be described in terms of normal ordered characteristic function
$C\left(\left\{ \beta_{j}\right\} \right)$ , it allows us to describe
quasidistribution of phases $\Theta\left(\left\{ \phi_{j}\right\} \right)$
directly in terms of characteristic function as \citep{perina2019quasidistribution}
\begin{equation}
\begin{array}{lcl}
\Theta\left(\left\{ \phi_{j}\right\} \right) & = & \frac{1}{\pi^{2n}}\mathcal{P}\int\Pi_{j=1}^{n}\frac{C\left(\left\{ \beta_{j}\right\} \right)}{\left(\beta_{j}\exp\left(-i\phi_{j}\right)-{\rm c.c.}\right)}d^{2}\beta_{j}\end{array},\label{eq:phaseQD}
\end{equation}
where $\mathcal{P}$ means the principal value of the integral, and
$\left\{ \beta_{j}\right\} =\left\{ \beta_{1},\beta_{2}\ldots,\beta_{n}\right\} $
are parameters of the characteristic function. Time evolution of the
normal ordered characteristic function in both weak and strong pump
cases can be described in the Gaussian form as 
\begin{equation}
\begin{array}{lcl}
C\left(\left\{ \beta_{j}\right\} ,t\right) & = & \left\langle \exp\left\{ \underset{j<k=\mathcal{S}}{\sum}\left[-\left(B_{j}\left(t\right)-\left|C_{j}\left(t\right)\right|\cos\eta_{j}\right)\left|\beta_{j}\right|^{2}+2\left|\beta_{j}\beta_{k}\right|\left(\left|D_{jk}\left(t\right)\right|\cos\epsilon_{jk}+\left|\bar{D}_{jk}\left(t\right)\right|\cos\bar{\epsilon}_{jk}\right)\right]\right\} \right\rangle \end{array}\label{eq:charF}
\end{equation}
in terms of quantum noise functions \citep{perina1991quantum} defined
as $B_{j}=\left\langle \Delta a_{j}^{\dagger}\Delta a_{j}\right\rangle ,\,C_{j}=\left\langle \left(\Delta a_{j}\right)^{2}\right\rangle ,\,D_{jk}=\left\langle \Delta a_{j}\Delta a_{k}\right\rangle ,$
and $\overline{D}_{jk}=-\left\langle \Delta a_{j}^{\dagger}\Delta a_{k}\right\rangle $
with $\left\{ \beta_{j}\right\} =\left\{ \beta_{\mathcal{S}}\right\} $.
Here, $\eta_{j}=\arg\left(C_{j}\right)$, $\epsilon_{jk}=\arg\left(D_{jk}\right)$,
and $\bar{\epsilon}_{jk}=\arg\left(\bar{D}_{jk}\right)$. The set
$\mathcal{S}$ is assumed ordered being $\mathcal{S}=\left\{ L,S,V,A\right\} $
and $\mathcal{S}=\left\{ S,V,A\right\} $ in case of weak and strong
pump cases, respectively. 

The single-mode nonclassicality is characterized by $s_{i}=B_{i}/\left|C_{i}\right|<1$.
This parameter $s_{i}>\cos\eta_{i}$ in turn also determines the bound
of the phase corresponding to nonclassical region \citep{perina2019quasidistribution}.
Similarly, we lack a classical description in two-mode nonclassical
region if $q_{jk}=\left(B_{j}-\left|C_{j}\right|\cos\eta_{j}\right)\left(B_{k}-\left|C_{k}\right|\cos\eta_{k}\right)/\left(\left|D_{jk}\right|+\left|\overline{D}_{jk}\right|\right)^{2}<1$.
This parameter also determines the corresponding threshold value of
phase in the nonclassical region \citep{perina2019quasidistribution}
as $1-q_{jk}<\sin^{2}\Psi_{jk}$, where $\Psi_{jk}=\epsilon_{jk}-\phi_{j}-\phi_{k}$.
The phases $\Psi_{jk}$ and $\bar{\Psi}_{jk}=\bar{\epsilon}_{jk}+\phi_{j}-\phi_{k}$
fulfill 
\[
\left|D_{jk}\right|^{2}\sin^{2}\Psi_{jk}+\left|\overline{D}_{jk}\right|^{2}\sin^{2}\bar{\Psi}_{jk}+2\left|D_{jk}\right|\left|\overline{D}_{jk}\right|\left(1-\cos\Psi_{jk}\cos\bar{\Psi}_{jk}\right)-\left(1-q_{jk}\right)\left(\left|D_{jk}\right|+\left|\overline{D}_{jk}\right|\right)^{2}>0.
\]
 The corresponding phase distributions are given in \citep{perina2019quasidistribution}
in a canonical form in dependence on $q_{jk}$.

\subsection{Raman process with weak pump}

Firstly, we begin with Raman process with weak pump conditions. The
perturbative analytic form of quantum noise terms in the characteristic
function (\ref{eq:charF}) in this case is reported in Appendix A.
We are interested in the off-resonant Raman process \citep{thapliyal2019nonclassicality}
where the dynamics is dependent upon two frequency detuning parameters,
particularly $\Delta\omega_{1}=(\omega_{S}+\omega_{V}-\omega_{L})$
and $\Delta\omega_{2}=(\omega_{L}+\omega_{V}-\omega_{A})$ as detuning
parameters in Stokes and anti-Stokes generation processes, respectively.
Here, $\omega_{L}$, $\omega_{S}$, $\omega_{A}$, and $\omega_{V}$
correspond to pump, Stokes, anti-Stokes, and phonon mode frequencies,
respectively. 

We have considered two limiting cases, $\Delta\omega_{1}=\Delta\omega_{2}=\delta_{1}$
(related to radiation modes) and $\Delta\omega_{1}=-\Delta\omega_{2}=\delta_{2}$
(related to vibrational mode), which we will refer to Case 1 and Case
2, respectively. In Case 1, the value of $s_{i}$ parameter for pump
and phonon modes are 

\begin{equation}
\begin{array}{lcl}
s_{L1} & = & \frac{2p\sin^{2}\frac{\delta_{1}t}{2}}{\Lambda_{1}}\sqrt{\frac{I_{A}}{I_{S}}}>\cos\eta_{L1},\\
s_{V1} & = & \frac{\left(I_{L}+p^{2}I_{A}\right)}{p\sqrt{I_{S}I_{A}}}>\cos\eta_{V1}.
\end{array}\label{eq:sin-non}
\end{equation}
Here and in what follows, $\Lambda_{i}^{2}=\left(\delta_{i}^{2}t^{2}+4\sin^{2}\frac{\delta_{i}t}{2}-2\delta_{i}t\sin\delta_{i}t\right)$,
intensity of each mode $I_{i}=\left|\xi_{i}\right|^{2}$, $p=\frac{\left|\chi\right|}{\left|g\right|}$,
where $g$ and $\chi$ are Stokes and anti-Stokes coupling constants,
respectively. Also, $\cos\eta_{L1}=-2\Re\left(\chi g^{*}\left\{ 2\sin^{2}\frac{\delta_{1}t}{2}+i\left(\delta_{1}t-\sin\delta_{1}t\right)\right\} \exp\left\{ i\left(\phi_{A}+\phi_{S}-2\omega_{L}t\right)\right\} \right)/\left(\left|\chi\right|\left|g\right|\Lambda_{1}\right)$.
Assuming coupling constant real and in the limits of zero detuning
(corresponding to resonant Raman process), $\eta_{L1}=-\phi_{S}-\phi_{A}+2\omega_{L}t$
is solely determined by the phase of Stokes and anti-Stokes modes
and pump frequency. Similarly, $\cos\eta_{V1}=-\Re\left(\chi g\exp\left\{ i\left(\phi_{A}-\phi_{S}-2\omega_{V}t+\delta_{1}t\right)\right\} \right)/\left(2\left|\chi\right|\left|g\right|\right)$
with $\omega_{V}$ as the phonon frequency, which reduces to $\eta_{V1}=\phi_{S}-\phi_{A}+2\omega_{V}t$
for real coupling constants for resonant Raman process. Similarly,
in Case 2, the same witness turns out to be 
\begin{equation}
\begin{array}{lcl}
s_{L2} & = & p\sqrt{\frac{I_{A}}{I_{S}}}>\cos\eta_{L2},\\
s_{V2} & = & \frac{2\sin^{2}\frac{\delta_{2}t}{2}\left(I_{L}+p^{2}I_{A}\right)}{p\Lambda_{2}\sqrt{I_{S}I_{A}}}>\cos\eta_{V2}.
\end{array}\label{eq:sin-non2}
\end{equation}
In this case, $\cos\eta_{L2}=-\Re\left(\chi g^{*}\exp\left\{ i\left(\phi_{S}+\phi_{A}-2\omega_{L}t-\delta_{2}t\right)\right\} \right)/\left(2\left|\chi\right|\left|g\right|\right)$
and $\cos\eta_{V2}=-2\Re\left(\chi g\left\{ 2\sin^{2}\frac{\delta_{2}t}{2}+i\left(\delta_{2}t-\sin\delta_{2}t\right)\right\} \exp\left\{ i\left(\phi_{A}-\phi_{S}-2\omega_{V}t\right)\right\} \right)/\left(\left|\chi\right|\left|g\right|\Lambda_{2}\right)$,
which are same but out of phase with $\eta_{L1}$ and $\eta_{V1}$
for real coupling constants in resonant Raman process, respectively.\textcolor{red}{{} }

\begin{figure}
\begin{centering}
\subfloat[]{\begin{centering}
\includegraphics[scale=0.5]{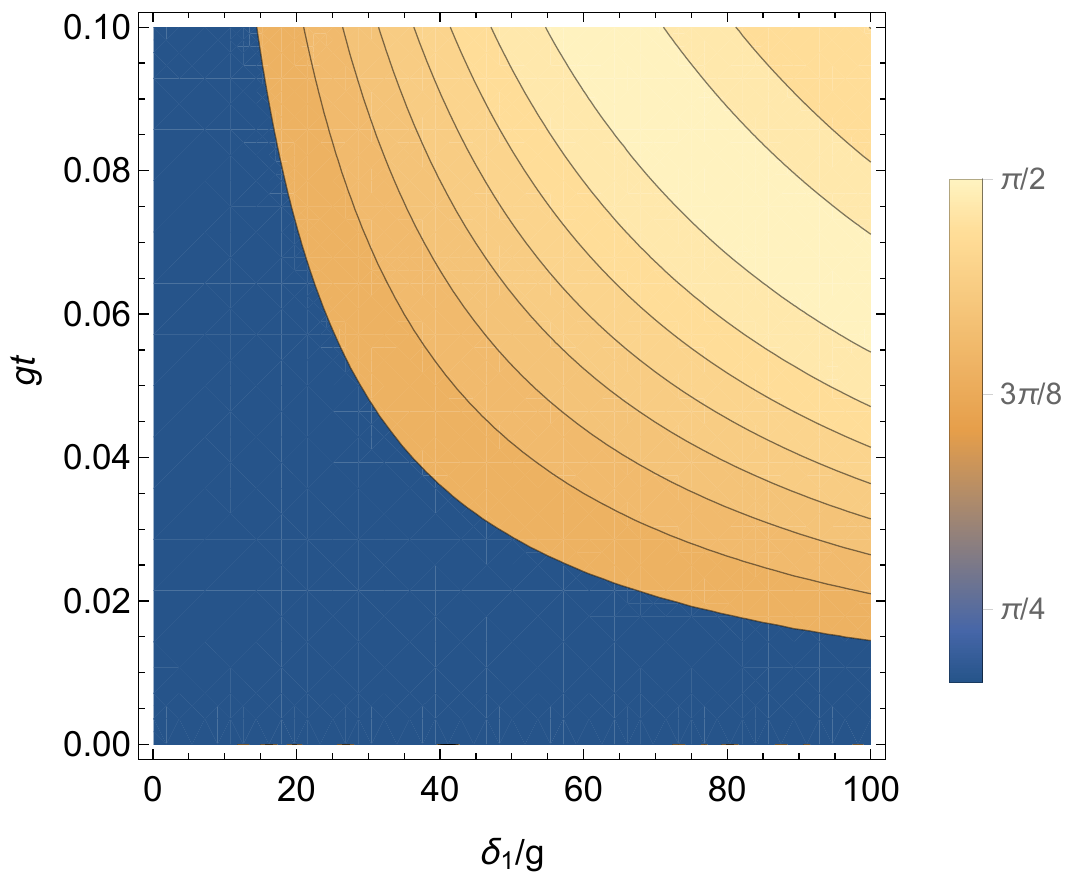} 
\par\end{centering}
} \subfloat[]{\begin{centering}
\includegraphics[scale=0.5]{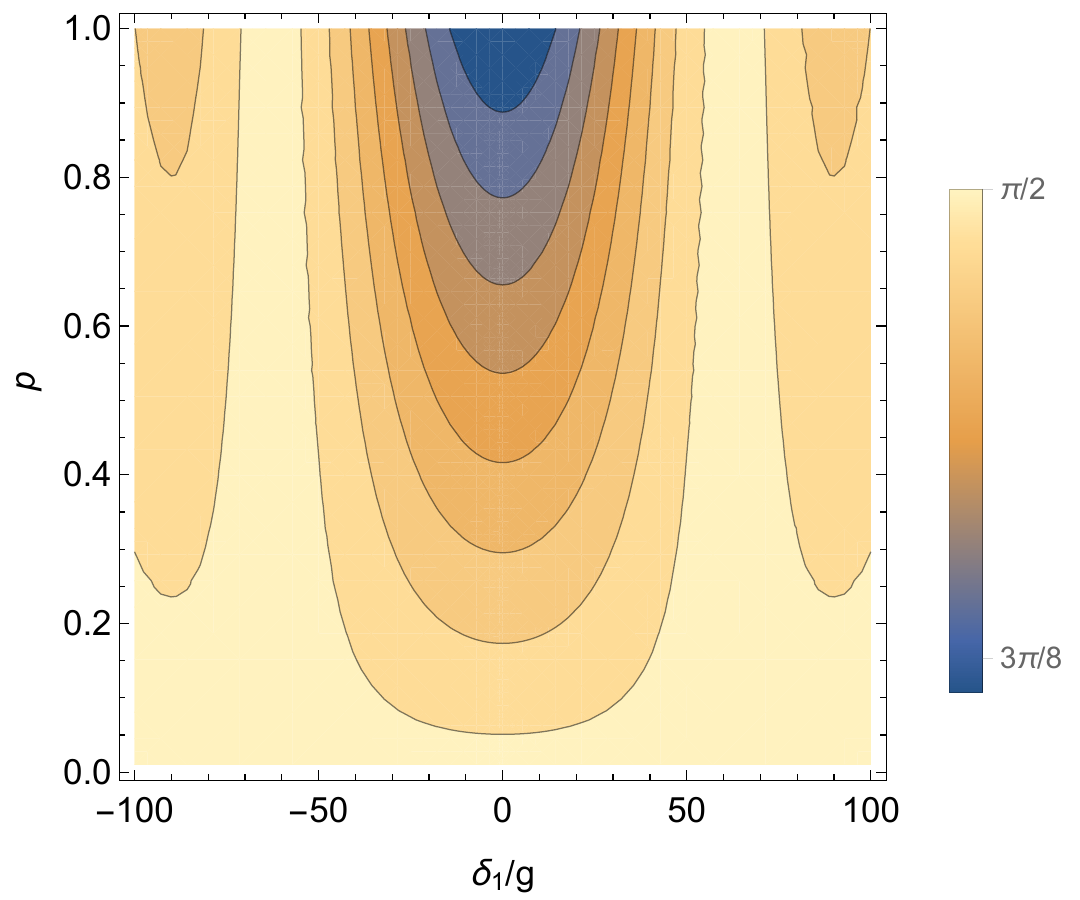} 
\par\end{centering}
} \subfloat[]{\begin{centering}
\includegraphics[scale=0.5]{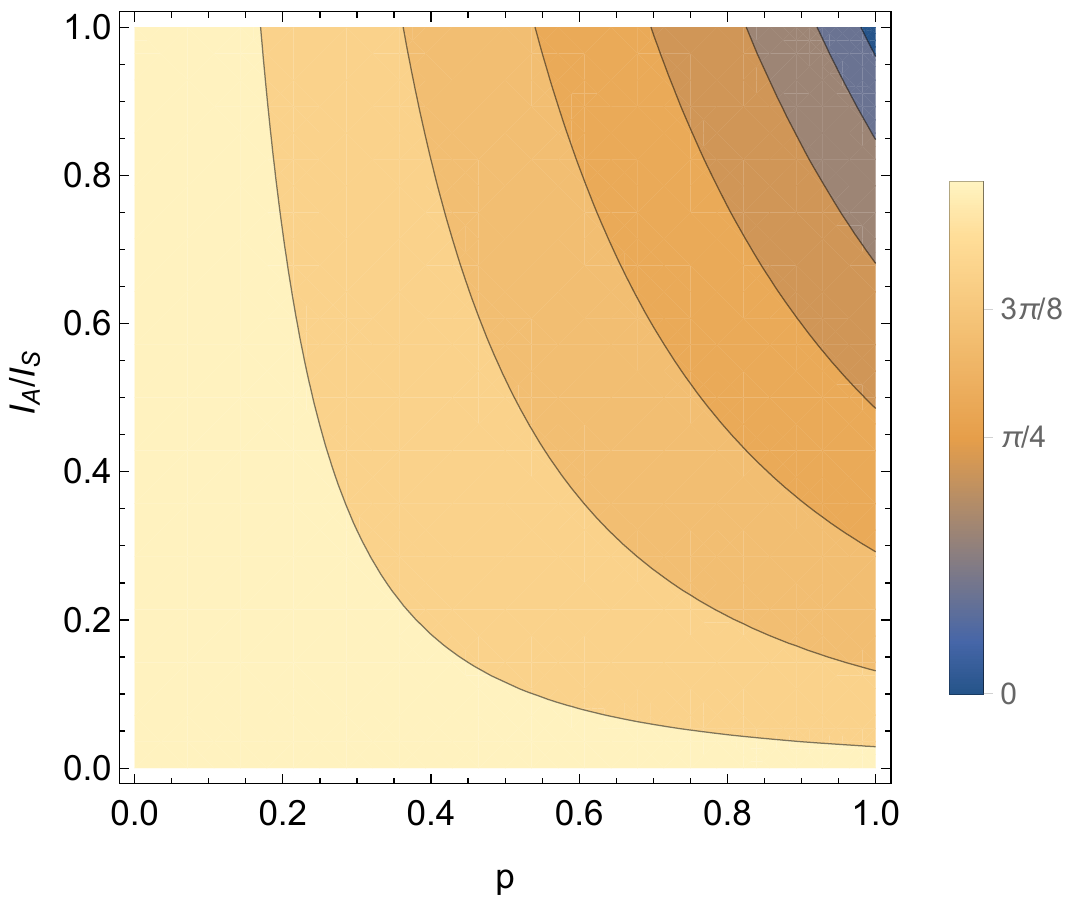}
\par\end{centering}
}
\par\end{centering}
\caption{\label{fig:sl} (Color online) Dependence of bound for $\eta_{L1}$
(which is given by the value of $\cos^{-1}s_{L1}$) on different parameters,
i.e., detuning parameter and (a) rescaled time and (b) the ratio of
anti-Stokes and Stokes coupling constants; (c) dependence of $s_{L2}$
on rescaled time and the ratio of anti-Stokes and Stokes coupling
constants. The plots are obtained by choosing $I_{S}=6$ and $I_{A}=1$
(wherever needed). All the quantities in this figure and rest of the
figures are dimensionless.}
\end{figure}
One can clearly verify that $\underset{\delta_{1}\rightarrow0}{\lim}s_{L1}=s_{L2}$
and $\underset{\delta_{2}\rightarrow0}{\lim}s_{V2}=s_{V1}$, which
represents results for Raman process at resonance {(using $\underset{\delta_{i}\rightarrow0}{\lim}2\sin^{2}\frac{\delta_{i}t}{2}/\Lambda_{i}=1$).
As the Cases 1 and 2 and small detuning fully determine the approximation with the ratios of coupling constants and intensities $\cos\eta_{L}<s_{L2}$
and $\cos\eta_{V}<s_{V1}$.}  Note that the obtained parameters
are independent of the initial phonon numbers. It is also easy to
check in case of spontaneous process that the parameters cannot be
defined for phonon mode. Further, it is safe to assume in case of
stimulated process that $\frac{I_{A}}{I_{S}}<<1$ and $p\approx1$,
and thus the sum of the phases of the Stokes and anti-Stokes modes
is described by a generalized quasiprobability distribution. However,
feasibility of describing the regular phase distribution for the difference
of the phases for resonant Raman process depends on $s_{V}=I_{L}/\left(p\sqrt{I_{S}I_{A}}\right)+s_{L}$,
which is dominated by the first term due to strong pump intensity,
and thus it shows classical behavior. Variation of obtained bound
of phase difference for pump mode in Case 1 for $\eta_{L1}=\cos^{-1}s_{L1}$
is shown in Fig. \ref{fig:sl}. Both time evolution and detuning parameter
enforce filtering of phase difference to allow a regular distribution
function in this case (Fig. \ref{fig:sl} (a)-(b)). In contrast, ratio
of anti-Stokes and Stokes coupling constants as well as the intensity,
however, diminish this filtering (Fig. \ref{fig:sl} (b)-(c)). 

\begin{figure}
\begin{centering}
\subfloat[]{\begin{centering}
\includegraphics[scale=0.6]{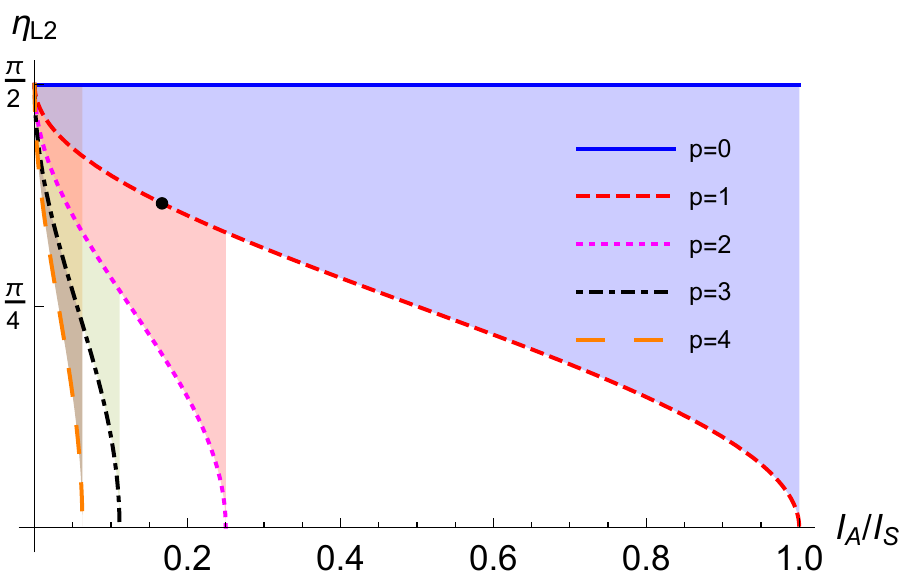}
\par\end{centering}
} \subfloat[]{\begin{centering}
\includegraphics[scale=0.6]{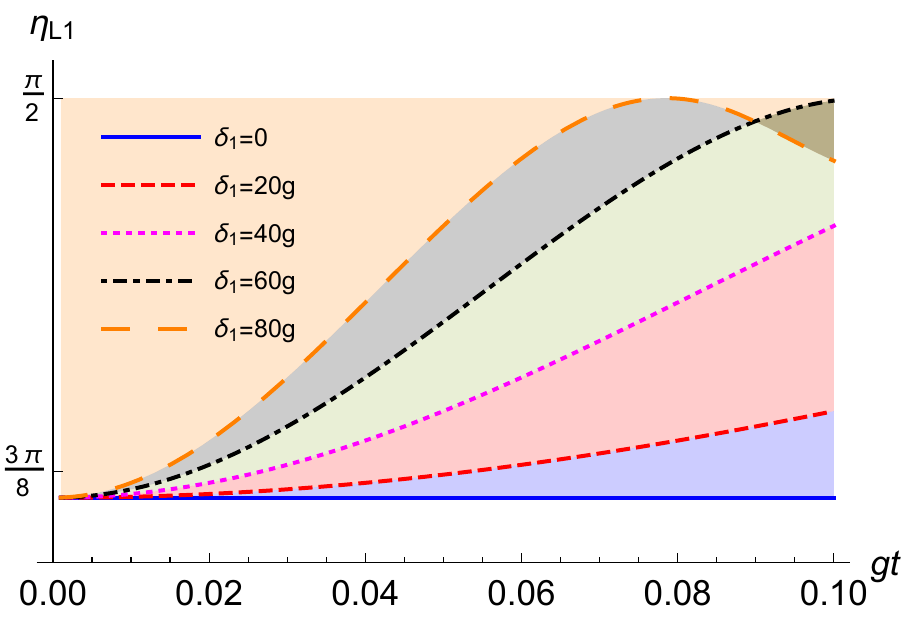} 
\par\end{centering}
} \subfloat[]{\begin{centering}
\includegraphics[scale=0.6]{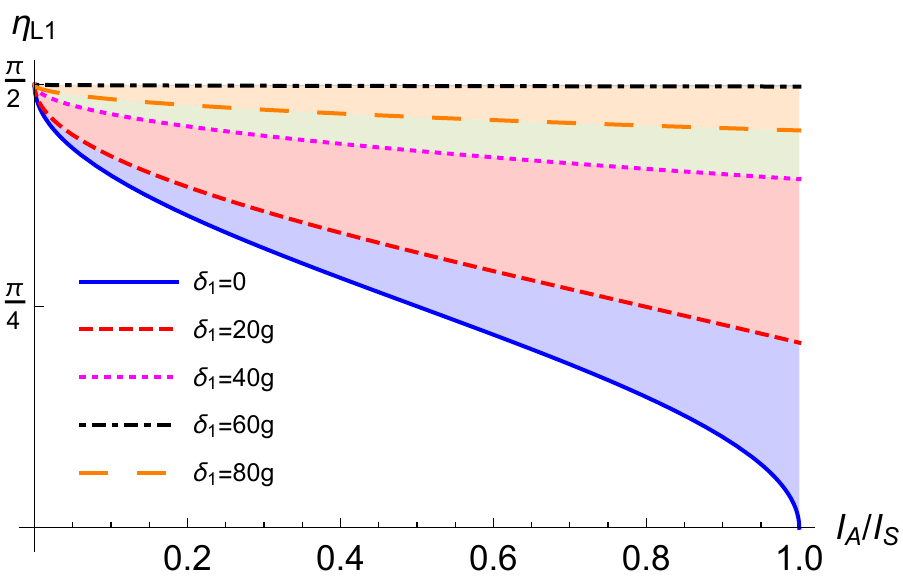}
\par\end{centering}
} 
\par\end{centering}
\caption{\label{fig:sl-2d}(Color online) (a) Dependence of $\eta_{L2}$ bound
(applicable for resonant Raman scattering, too) on rescaled time and
$p$. Dependence of $s_{L1}$ on (b) rescaled time and (c) the ratio
of anti-Stokes and Stokes intensities for different values of detuning
parameter. The allowed values of phase difference are shown as the
shaded region. We have chosen $I_{S}=6$ and $p=1$. For the point
marked in (a) the effect of frequency detuning parameter is illustrated
in (b), while similar variation for the corresponding line is shown
in (c). }
\end{figure}

The bound of phase difference in case of pump mode in Case 2 is also
applicable to resonant Raman scattering. Variation of bound for $\eta_{L2}$
with the ratio of anti-Stokes and Stokes coupling constants (for different
Raman active materials \citep{parra2016stokes}) as well as intensity
is shown in Fig. \ref{fig:sl-2d} (a). We have marked a point on Fig.
\ref{fig:sl-2d} (a) and shown its time evolution for off-resonant
Raman process for $\eta_{L1}$ in Fig. \ref{fig:sl-2d} (b). The similar
variation of the line containing the point with the ratio of anti-Stokes
and Stokes intensity and frequency detuning is shown in Fig. \ref{fig:sl-2d}
(c). Both increasing time evolution and frequency detuning require
more filtering, while $\frac{I_{A}}{I_{S}}$ has the opposite effect. 

\begin{figure}
\begin{centering}
\includegraphics[scale=0.9]{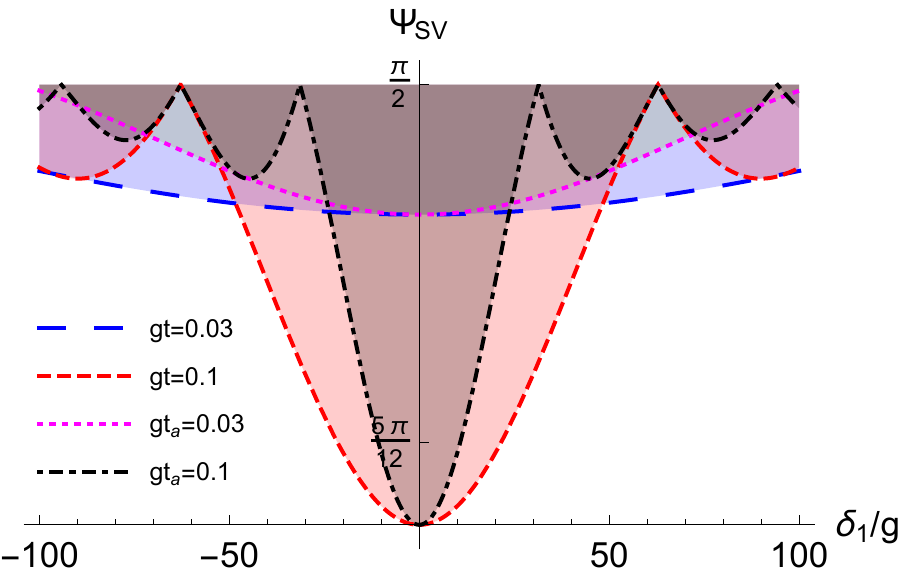}
\par\end{centering}
\caption{\label{fig:app}(Color online) Variation of complete analytic solution
and simplified approximate solution (with a in the subscript in the
plot legends) for $\Psi_{SV}$ parameters with frequency detuning.
The allowed values of phase are shown as the shaded region. The results
are obtained for spontaneous case with $I_{L}=10$ and $p=1$. Similar
behavior was observed for $\Psi_{LV}$ in the stimulated case.}
\end{figure}

In the present case, $q_{ij}$ parameter based on two-mode nonclassicality
is used further to study filtering of phase difference in such cases.
A much simpler form is $q_{SV}\approx\left|g\right|^{2}t^{2}I_{L}{\rm sinc}^{2}\delta_{i}t$
which can be obtained in the spontaneous case. The simplified form
is less than 1 imposing condition for short-time limit \citep{thapliyal2019nonclassicality}
$\left|g\right|t\sqrt{I_{L}}<1$ and assuming $\delta_{i}t\ll1$.
Thus, $q_{SV}$ can be observed in good agreement with the complete
expression in the short-time and small detuning limits as shown in Fig.
\ref{fig:app}, where we have exhibited the allowed region for phase
$\Psi_{SV}$ in different cases by different colors. Due to the complex
structure of bound for the phase difference parameter in off-resonant
case, we report here corresponding parameter for zero detuning as
$\Psi_{SV}=\arg\left[-gte^{-it\left(\omega_{b}+\omega_{c}\right)}\left(g^{*}t\sqrt{I_{S}I_{V}}e^{i\left(\phi_{S}+\phi_{V}\right)}+2\chi t\sqrt{I_{A}I_{V}}e^{i\left(\phi_{A}-\phi_{V}\right)}-2i\sqrt{I_{L}}e^{i\phi_{L}}\right)\right]$.
The simplified approximate solution is marked with 'a' in the subscript in Fig.
\ref{fig:app}.
Clearly, the approximate solution always gives the narrower region
than that is shown by the complete analytic solution. Similarly, in
the partial stimulated case (i.e., $I_{A}\neq0=I_{S}$), $q_{LV}\approx\left|g\right|^{2}\left(t^{2}I_{L}+p^{2}t^{2}I_{A}\right){\rm sinc}^{2}\delta_{i}t$,
which qualitatively concludes the same as for $\Psi_{SV}$ in Fig.
\ref{fig:app}. The corresponding behavior for $\Psi_{LV}=\arg\left[-e^{-it\left(\omega_{V}+\omega_{L}\right)}\left(\left\{ \left|g\right|^{2}+\left|\chi\right|^{2}\right\} t^{2}\sqrt{I_{L}I_{V}}e^{i\left(\phi_{L}+\phi_{V}\right)}-2i\chi t\sqrt{I_{A}}e^{i\phi_{A}}\right)\right]$
was obtained for resonant Raman stimulated process. 

While for pump-Stokes case, we have obtained in partial stimulated
cases that $q_{LS}=1$ if $I_{A}\neq0=I_{S}$, it is $q_{LS}=0$ for
$I_{S}\neq0=I_{A}$. This means that the compound pump-Stokes mode
is classical in the former case, coherence from the laser mode is
directly transferred to the Stokes mode and the wave distribution
is proportional to the Dirac $\delta$-function, as $B_{L}B_{S}-\left|\bar{D}_{LS}\right|^{2}=0$
and $C_{L}=C_{S}=D_{LS}=0$. In the latter case, there is no filtering
of phase differences as $B_{L}=C_{L}=0$ and $D_{LS}\neq0\neq B_{S}$.
Here, $\sin\Psi_{LS}=-2\Re\left(\left\{ 2\sin^{2}\frac{\delta_{j}t}{2}+i\left(\delta_{j}t-\sin\delta_{j}t\right)\right\} \exp\left\{ -i\left(\phi_{S}+\phi_{L}-\left(\omega_{L}+\omega_{S}\right)t\right)\right\} \right)/\Lambda_{j}$
in both Case 1 and Case 2. Similarly, one can calculate $\sin\bar{\Psi}_{LS}=\Re\left(\chi g^{*}\exp\left\{ i\left(\phi_{A}-\phi_{L}-\left(\omega_{L}-\omega_{S}\right)t\right)\right\} \right)/\left(\left|\chi\right|\left|g\right|\right)$
and $\sin\bar{\Psi}_{LS}=\Re\left(\chi g^{*}\exp\left\{ i\left(\phi_{A}-\phi_{L}-\left(\omega_{L}-\omega_{S}-\delta_{2}\right)t\right)\right\} \right)/\left(\left|\chi\right|\left|g\right|\right)$
in Cases 1 and 2, respectively, to verify phase relations.

\begin{figure}
\begin{centering}
\includegraphics[scale=0.9]{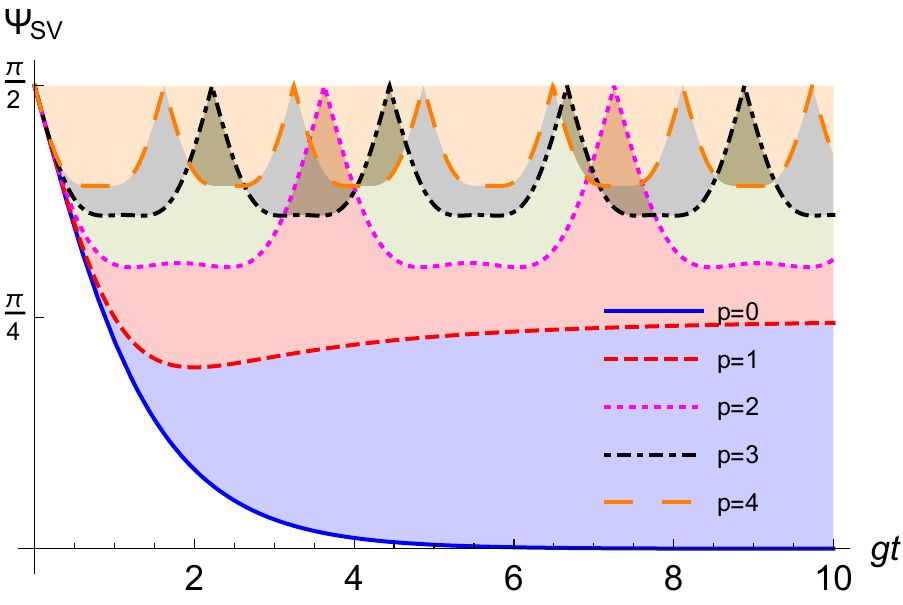}
\par\end{centering}
\caption{\label{fig:str} (Color online) Time evolution of $\Psi_{SV}$ parameter
for different values of the ratio of anti-Stokes and Stokes coupling
constants. This parameter filters the value of $\Psi_{SV}=-\left(\Phi_{L}-\omega_{V}t\right)$
(shown as the shaded region). Note that the quantity shown in the
plot coincides with $\Psi_{SA}=-\left(2\Phi_{L}-\omega_{A}t\right)$. }
\end{figure}

\subsection{Raman process with strong pump}

An exact closed form analytic solution and characteristic function
\citep{pieczonkova1981statistical} can be obtained for strong pump
resonant Raman process. The solution and its brief detail are given
in Appendix B. In this section, we study the phase properties of the
Raman process with strong coherent classical pump. Specifically, the
filtering in phase distributions cannot be studied from the signatures
of single-mode nonclassicality criterion as all noise parameters $C_{i}=0$.
Thus, using two-mode criterion we obtained 
\begin{equation}
\begin{array}{lcl}
q_{SV} & = & q_{SA}=\frac{\left(\left(p^{2}-1\right)\sin^{2}\left(gt\sqrt{p^{2}-1}\right)+2p^{2}\sin^{4}\left(\frac{gt\sqrt{p^{2}-1}}{2}\right)\right)}{\left(p^{2}-\cos\left(gt\sqrt{p^{2}-1}\right)\right)^{2}},\\
q_{VA} & = & 1.
\end{array}\label{eq:two-non}
\end{equation}
Here, $\sin\Psi_{SV}=-\Im\left(g\right)\cos\left(\Phi_{L}-\omega_{V}t\right)+\Re\left(g\right)\sin\left(\Phi_{L}-\omega_{V}t\right)/\left|g\right|$
with $\Phi_{L}$ as the phase of the strong classical pump beam, which
becomes $\Psi_{SV}=-\Phi_{L}+\omega_{V}t$ for real coupling constants.
Similarly, $\sin\Psi_{SA}=-\Im\left(\chi g\exp\left\{ 2i\Phi_{L}-i\omega_{A}t\right\} \right)/\left(\left|\chi\right|\left|g\right|\right)$
which becomes $\Psi_{SA}=-2\Phi_{L}+\omega_{A}t$ for real coupling
constants.

Similarly, it is possible to calculate $1-q_{jk}<\sin^{2}\bar{\Psi}_{jk}$
as $\sin\bar{\Psi}_{VA}=-\Im\left(\chi^{*}\exp\left\{ -i\Phi_{L}+i\left(\omega_{A}-\omega_{V}\right)t\right\} \right)/\left|\chi\right|$,
which becomes $\bar{\Psi}_{VA}=\Phi_{L}-\left(\omega_{A}-\omega_{V}\right)t$
for real coupling constants. As $q_{VA}=1$ in this case, all the
values of $\bar{\Psi}_{VA}$ are allowed. Note that although we are
using the same coupling constant parameters in both weak and strong
pump cases, pump amplitude is originally included in the coupling
amplitudes (i.e., in the latter, these constants will be $\sqrt{I_{L}}$
times the corresponding constants in the former case). For classical
strong pumping, the compound anti-Stokes-phonon mode corresponds with
the process of frequency-conversion conserving coherence $\left(B_{A}B_{V}-\left|\bar{D}_{VA}\right|^{2}=0\right)$
and the wave distribution is proportional to the Dirac $\delta$-function.
In short interaction times there is no quantum noise in the anti-Stokes
mode, which has a tendency to conserve coherence, being attenuated.
However, for strong pumping and large interaction times also in the
anti-Stokes mode the quantum noise is developed. 

In Fig. \ref{fig:str}, time evolution of $\Psi_{SV}$ (and thus $\Psi_{SA}$,
too) causes filtering of the phase of the pump mode with respect to the
frequency of the phonon mode and shows that for different values of
$p$ the phase distribution becomes singular at different rescaled
times. Further, from the figure it can be seen that with increase
in the anti-Stokes coupling with respect to Stokes coupling allowed
range of values of $\Psi_{SV}$ (and similarly $\Psi_{SA}$) decreases.
For instance, in case of zero anti-Stokes coupling (which can be assumed
in the weak pump conditions in the present case) there is no filtering
observed after certain value of rescaled time. In contrast, with higher
values of anti-Stokes coupling with respect to Stokes coupling the
pump phases are filtered to a smaller range of values around $\frac{\pi}{2}$. 

\section{Conclusions \label{sec:Conclusion}}

Quasidistribution of phases for the Raman process is obtained from
corresponding Glauber-Sudarshan $P\left(\left\{ \alpha_{j}\right\} \right)$
quasidistribution function by integrating over moduli of complex amplitudes.
Thus obtained unnormalized quasidistribution of phases contains signatures
of nonclassicality. Here, we have studied the effect of these nonclassical
behavior in the Raman process under both weak and strong pump conditions
on corresponding phase properties. The weak pump case is studied by
performing complete quantum treatment to obtain perturbative solution
which is applicable to both resonant and off-resonant Raman processes.
In contrast, with parametric approximation, i.e., strong classical
pump, an exact solution is possible, which is obtained only for resonant
Raman process. 

The single-mode nonclassicality is observed only with a complete quantum
treatment, and the phase differences of phases described by single
nonclassical modes are required to be filtered to describe a regular
distribution function. The compound anti-Stokes-phonon mode is observed
to be classical and coherence is conserved in strong pump case and
corresponding wave distribution can be described by Dirac $\delta$-functions.
Similar behavior is observed in special case in pump-Stokes mode with
weak pump. Compound Stokes-phonon mode shows nonclassical features
of phases with both weak and strong pumping, which behavior is similar
to pump-phonon (Stokes-anti-Stokes) mode with weak (strong) pump. 

The present study shows the effect of nonclassicality present in the
output modes of the Raman process on corresponding phase properties
and will be helpful in understanding the behavior observed in the
experiments. In particular, in \citep{armata2016quantum}, the authors
illustrated possibility to distinguish classical and quantum phases
in optomechanical interference experiment with weak coupling and for
small photon numbers and low temperature. Our suggestion of the interference
experiment \citep{perina2019quasidistribution} for obtaining the
presented nonclassical phase effects has to follow such conditions.

\textbf{Acknowledgement:} {{} Authors acknowledge the
financial support from the Operational Programme Research, Development
and Education - European Regional Development Fund project no. CZ.02.1.01/0.0/0.0/16
019/0000754 of the Ministry of Education, Youth and Sports of the
Czech Republic.}

\bibliographystyle{apsrev}
\bibliography{hyRa}

\section*{Appendix A: Finite time coefficient of characteristic function}

\setcounter{equation}{0} 
\global\long\def\theequation{A.\arabic{equation}}%

In the present case, the obtained quantum noise function terms are
as follows \citep{thapliyal2019nonclassicality}
\begin{equation}
\begin{array}{lcl}
B_{L}\left(t\right) & = & \frac{2\left|\chi\right|^{2}\left|\xi_{A}\right|^{2}\left(1-\cos\delta_{i}t\right)}{\delta_{i}^{2}},\\
B_{S}\left(t\right) & = & \frac{2\left|g\right|^{2}\left|\xi_{L}\right|^{2}\left(1-\cos\delta_{i}t\right)}{\delta_{i}^{2}},\\
B_{V}\left(t\right) & = & B_{L}\left(t\right)+B_{S}\left(t\right),\\
C_{L}\left(t\right) & = & \frac{2\xi_{S}\xi_{A}\chi g^{*}e^{-it\left(\delta_{i}+2\omega_{L}\right)}\left(\pm2\sin^{2}\left(\frac{\delta_{i}t}{2}\right)-i\left(\sin\delta_{1}t-\delta_{1}te^{i\delta_{1}t}\right)\right)}{\delta_{i}^{2}},\\
C_{V}\left(t\right) & = & \frac{2\xi_{S}^{*}\xi_{A}\chi ge^{it\left(\delta_{i}-2\omega_{V}\right)}\left(\mp2\sin^{2}\left(\frac{\delta_{i}t}{2}\right)+i\left(\sin\delta_{2}t-\delta_{2}te^{-i\delta_{2}t}\right)\right)}{\delta_{i}^{2}},\\
D_{LS}\left(t\right) & = & \frac{\xi_{S}\xi_{L}\left|g\right|^{2}e^{-it\left(\omega_{L}+\omega_{S}\right)}\left(-2\sin^{2}\left(\frac{\delta_{i}t}{2}\right)-i\left(\sin\delta_{i}t-\delta_{i}t\right)\right)}{\delta_{i}^{2}},\\
D_{LV}\left(t\right) & = & \frac{2i\chi\xi_{A}\sin\left(\frac{\delta_{i}t}{2}\right)e^{-it\left(\omega_{L}+\omega_{V}\pm\frac{\delta_{i}}{2}\right)}}{\delta_{i}}-\frac{\xi_{L}\xi_{V}e^{-it\left(\omega_{L}+\omega_{V}\right)}\left(\left(\left|g\right|^{2}+\left|\chi\right|^{2}\right)\left\{ 2\sin^{2}\left(\frac{\delta_{i}t}{2}\right)+i\left(\sin\delta_{i}t-\delta_{i}t\right)\right\} -2i\left|\chi\right|^{2}\left(\sin\delta_{1}t-\delta_{1}t\right)\right)}{\delta_{i}^{2}},\\
D_{LA}\left(t\right) & = & \frac{\xi_{L}\xi_{A}\left|\chi\right|^{2}e^{-it\left(\omega_{L}+\omega_{A}\right)}\left(-2\sin^{2}\left(\frac{\delta_{i}t}{2}\right)\mp i\left(\sin\delta_{i}t-\delta_{i}t\right)\right)}{\delta_{i}^{2}},\\
D_{SV}\left(t\right) & = & \frac{2i\chi\xi_{L}\sin\left(\frac{\delta_{i}t}{2}\right)e^{-it\left(\frac{\delta_{i}}{2}+\omega_{S}+\omega_{V}\right)}}{\delta_{i}}+\frac{\xi_{S}\xi_{V}\left|g\right|^{2}e^{-it\left(\omega_{S}+\omega_{V}\right)}\left(-2\sin^{2}\left(\frac{\delta_{i}t}{2}\right)+i\left(\sin\delta_{i}t-\delta_{i}t\right)\right)}{\delta_{i}^{2}}\\
 & + & \frac{2\xi_{V}^{*}\xi_{A}\chi g\left(\mp2\sin^{2}\left(\frac{\delta_{i}t}{2}\right)+i\left(\sin\delta_{2}t-\delta_{2}te^{-i\delta_{2}t}\right)\right)e^{-it\left(\omega_{S}+\omega_{V}-\delta_{i}\right)}}{\delta_{i}^{2}},\\
D_{SA}\left(t\right) & = & \frac{\xi_{L}^{2}\chi^{*}ge^{-it\left(\omega_{S}+\omega_{A}\right)}\left(-2\sin^{2}\left(\frac{\delta_{1}t}{2}\right)-4\sin^{2}\left(\frac{\delta_{2}t}{2}\right)e^{i\delta_{2}t}-i\left(\sin\delta_{1}t-\delta_{1}t\right)\right)}{\delta_{i}^{2}},\\
D_{VA}\left(t\right) & = & \frac{\xi_{V}\xi_{A}\left|\chi\right|^{2}e^{-it\left(\omega_{V}+\omega_{A}\right)}\left(-2\sin^{2}\left(\frac{\delta_{i}t}{2}\right)\mp i\left(\sin\delta_{i}t-\delta_{i}t\right)\right)}{\delta_{i}^{2}},\\
\overline{D}_{LS}\left(t\right) & = & \frac{4\chi^{*}g\xi_{L}\xi_{A}^{*}\sin^{2}\left(\frac{\delta_{i}t}{2}\right)e^{-it\left(\delta_{2}-\omega_{L}+\omega_{S}\right)}}{\delta_{i}^{2}}.
\end{array}\label{eq:Terms-in-CF}
\end{equation}
The rest of the terms in Eq. (\ref{eq:charF}) are zero. Here, $\delta_{1}\neq0$
and $\delta_{2}=0$ ($\delta_{1}=0$ and $\delta_{2}\neq0$) correspond
to Case 1 (Case 2). 

\section*{Appendix B: Strong-pump solution (some features)}

\setcounter{equation}{0} 
\global\long\def\theequation{B.\arabic{equation}}%

Quantum noise fluctuation terms in the characteristic function can
be written as \citep{pieczonkova1981statistical}
\begin{equation}
\begin{array}{lcl}
B_{V} & = & \frac{\left|g\right|^{2}\sin^{2}\left(t\sqrt{\left|\chi\right|^{2}-\left|g\right|^{2}}\right)}{\left(\left|\chi\right|^{2}-\left|g\right|^{2}\right)},\\
B_{A} & = & \frac{\left|\chi\right|^{2}\left|g\right|^{2}\left(1-\cos\left(t\sqrt{\left|\chi\right|^{2}-\left|g\right|^{2}}\right)\right)^{2}}{\left(\left|\chi\right|^{2}-\left|g\right|^{2}\right)^{2}},\\
B_{S} & = & B_{V}+B_{A},\\
D_{SA} & = & \frac{-\chi g\left(1-\cos\left(t\sqrt{\left|\chi\right|^{2}-\left|g\right|^{2}}\right)\right)\left\{ \left|\chi\right|^{2}-\left|g\right|^{2}\cos\left(t\sqrt{\left|\chi\right|^{2}-\left|g\right|^{2}}\right)\right\} \exp\left(2i\Phi_{L}-i\left(\omega_{A}+\omega_{S}\right)t\right)}{\left(\left|\chi\right|^{2}-\left|g\right|^{2}\right)^{2}},\\
D_{VS} & = & \frac{ig\sin\left(t\sqrt{\left|\chi\right|^{2}-\left|g\right|^{2}}\right)\left\{ \left|\chi\right|^{2}-\left|g\right|^{2}\cos\left(t\sqrt{\left|\chi\right|^{2}-\left|g\right|^{2}}\right)\right\} \exp\left(i\Phi_{L}-i\left(\omega_{V}+\omega_{S}\right)t\right)}{\left(\left|\chi\right|^{2}-\left|g\right|^{2}\right)^{3/2}},\\
\bar{D}_{VA} & = & \frac{i\left|g\right|^{2}\chi^{*}\sin\left(t\sqrt{\left|\chi\right|^{2}-\left|g\right|^{2}}\right)\left(1-\cos\left(t\sqrt{\left|\chi\right|^{2}-\left|g\right|^{2}}\right)\right)\exp\left(-i\Phi_{L}-i\left(\omega_{V}-\omega_{A}\right)t\right)}{\left(\left|\chi\right|^{2}-\left|g\right|^{2}\right)^{3/2}},
\end{array}\label{eq:noise}
\end{equation}
while the rest of the terms are zero.
\end{document}